\documentclass[reprint,groupedaddress,superscriptaddress,floatfix,tightenlines,aps]{revtex4-2}
\usepackage[utf8x]{inputenc}
\usepackage[english]{babel}
\usepackage[T1]{fontenc}
\usepackage{lmodern}
\usepackage{dsfont}

\usepackage[margin=0.6in]{geometry}
\usepackage{fancyhdr}
\usepackage{amsmath,amsfonts,amssymb,amsthm,bm,times}
\usepackage{graphicx,color}
\usepackage[caption=false]{subfig}
\usepackage{textcomp}
\usepackage{lipsum}
\usepackage{verbatim}
\usepackage{blindtext}
\usepackage{makecell}
\usepackage{ulem}
\usepackage{booktabs}
\usepackage[toc,page]{appendix}

\usepackage{microtype}

\usepackage{braket}

\begin{document}

\title{
Valley-dependent tunneling 
through electrostatically created quantum dots\\ 
in heterostructures of graphene with hexagonal boron nitride 
}

\author{A. Belayadi}
\email{abelayadi@usthb.dz}
\affiliation{University Of Science And Technology Houari Boumediene, Bab-Ezzouar, Algeria.}
\author{N. A. Hadadi}
\email{0naif099@gmail.com}
\affiliation{King Fahd University of Petroleum and Minerals,  Saudi Arabia.}
\author{P. Vasilopoulos }
\email{p.vasilopoulos@concordia.ca}
\affiliation{Department of Physics, Concordia University, 7141 Sherbrooke Ouest, Montr\'eal, Qu\'ebec H4B 1R6, Canada.}
\author{A. Abbout}
\email{adel.abbout@kfupm.edu.sa}
\affiliation{King Fahd University of Petroleum and Minerals,  Saudi Arabia.}

\begin{abstract}
Kelvin probe force microscopy (KPFM) has been employed to probe  charge carriers in a graphene/hexagonal boron nitride (hBN) heterostructure [
Nano Lett,  {\bf 21}, 5013 ( 2021)]. 
We propose an  approach for operating valley filtering based on  the KPFM-induced potential  $U_0$ instead of using 
external or induced pseudo-magnetic fields in strained graphene. Employing a tight-binding model, we investigate the parameters and rules leading to valley filtering in the presence of 
a graphene quantum dot (GQD) created by  the KPFM tip. 
This 
model leads to a resolution of different transport channels in  reciprocal space, where the electron transmission probability at each Dirac cone (${\bf {\bf K_1=-K}}$ and ${\bf {\bf K_2=+K}}$) is evaluated separately. The results show that 
$U_0$ and the Fermi energy $E_F$ control (or invert) the valley polarization, if electrons are allowed to flow through a given valley. The resulting  valley filtering is allowed only if the signs of $E_F$ and $U_0$ 
are the same.  If they are different, the valley filtering is destroyed and might occur only at some resonant states affected by $U_0$. 
Additionally,  there are  
independent valley modes characterizing the conductance oscillations near the vicinity of the resonances, whose strength  increases 
with $U_0$ 
and  are similar to those ocurring in resonant tunneling in quantum antidots and to the Fabry-Perot oscillations.
Using KPFM, to probe the charge carriers, and graphene-based structures to control  valley transport,  provides an efficient way for attaining valley filtering without involving external or pseudo-magnetic fields as in previous proposals.

\end{abstract}

\maketitle

\section{introduction}\label{sec1}
Graphene-based materials are 
excellent candidates for spintronic applications. Indeed, the presence of one or several types of spin-orbit couplings (SOCs) \cite{fhck, wst, kzaw, zfghg} led to many experimental and theoretical studies of these materials in order to control spin-transport properties 
in ultra thin spintronic devices \cite{zfs, belaya}. Besides  potential use in spintronics, many recent applications have adopted graphene as an essential material to constitute unique (fundamental) platforms in valleytronics \cite{costa, lltoa, yi-wen, d-zamb}. In this context, investigating valley filtering in  graphene-based devices 
may facilitate the 
 use of the valley degree of freedom in ${\bf k}$ space, instead of the spin degree of freedom, as an alternative basis for future applications in valleytronics.
 
Previous 
valley-filtering proposals have used a graphene layer, with uniform zigzag edges, and stressed it in a particular way that leads to the emergence of pseudo-magnetic fields (PMFs) \cite{deJuan, settnes, Milovan, Stegmann}. It has also been shown that the valley-filtering process might occur in a honeycomb lattice that contains a line of heptagon-pentagon defects \cite{Gunlycke1, Gunlycke2, YLiu}. Further, recent scanning tunneling microscopy (STM) and Kelvin probe force microscopy (KPFM) experiments claimed that by breaking the potential symmetry in the substrates of graphene-based heterostructures, by applying real magnetic fields, the valley degeneracy might be lifted if some conditions are fulfilled \cite{Orlof, Freitag, Wyatt, SYLi}. Therefore, any valley polarization might be measured through valley-split Landau levels (LLs) \cite{Freitag, zfghg, Giavaras1, Giavaras2, YSu, Chizhova} instead of PMFs.

Very recently, has been observed 
a nanoscale valley splitting 
has been observed in  confined states of  graphene quantum dots. In this case, the presence of a magnetic field and an STM-induced potential, originating from the boron nitride substrate beneath the graphene layer, provide an alternative device for valleytronics  \cite{ Freitag, Wyatt, SYLi}. However, in such cases the STM tip  breaks the electron-hole symmetry
and the magnetic field breaks the time-reversal symmetry; this will lead to an interplay between spintronics and valleytronics. 

The question then arises 
whether an alternative way exists to lift the valley degeneracy without confinement, that traps electrons around the STM potential, and without lifting the spin degeneracy. Indeed, from an application point of view and for a better tunability of transport properties, one needs to avoid the confinement of the electrons by the STM-induced potential since   they could tunnel through the induced potential barrier and contribute to the transmitted charge  or valley current. 
Fortunately,  several works have shown that the presence of a magnetic field $B$ along with the STM-induced potential do not always favor the confinement. For instance, in the case of a Gaussian shaped STM 
potential 
in a weak  field $B$, electrons are more likely to escape into the induced potential barrier 
 \cite{Orlof}. More precisely, in a weak  field $B$ with a circularly symmetric potential portrayed by a Gaussian model \cite{Orlof, SYLi, Giavaras2, Maksym}, the confinement  leads to a 
 compromise between the strengths of the potential and  of the field $B$.

As current conservation 
between the source and the drain leads of the graphene flake is desired,   a STM tip is not well suited for probing charge carriers since the current 
from the source reservoir will tunnel through the  tip as well. 
We need  an alternative method to create a graphene quantum dot (GQD) and keep the current  conserved.
 Fortunately, KPFM has been recently adopted as an efficient method in which tunneling can be neglected  \cite{Samaddar}. In contrast to STM, KPFM can induce an electrostatic potential and form a GQD  on a surface without 
the effects of local tip-gating. Indeed,  this is so because it is performed at slightly larger tip-sample distances, such that tunneling and van der Waals forces are significantly minimized \cite{Wyatt, Samaddar}.  

Based on the  arguments stated above and in order to better focus on valley polarization in graphene/hBN heterostructures with induced quantum dots, with the electron transmission probability  accounted for ${\bf K_1}$ and ${\bf K_2}$ independently, it is strongly recommended to avoid both confinement and tunneling of electrons as well as lifting the spin degeneracy caused by a magnetic field.  Accordingly, we investigate the valley polarized conductance in a graphene monolayer placed on top of a hBN substrate, with a voltage  induced by a KPFM tip, in the absence of  a  magnetic field. 

The results are  organized as follows. In Sec. \ref{sec2} we describe the 
graphene/hBN heterostructure in the presence of a quantum dot created electrostatically by KPFM.  
We then use a tight-binding model  to investigate valley-dependent transport. 
In Sec. \ref{sec3} we present and discuss  numerical results and  in
 Sec. \ref{sec4} a summary. 

\section{Model and Methods}\label{sec2}

We consider a  graphene/hBN heterostructure as shown in Fig. \ref{fig1}(a). 
A charge current at the graphene surface is controlled by the bias voltage $V_B$  applied between the source (S) and drain (D) 
 leads. The KPFM tip acts as a top gate $V_T$ and tunes the potential, which induces an electric field that forms a stationary distribution (see Fig. \ref{fig1} (c)) of the charges on the hBN substrate \cite{Orlof,  SYLi,  Wyatt}.  To evaluate the resulting 
 screened potential $U$ 
 several authors have solved the Poisson equation 
 self-consistently assuming a 
 KPFM-induced voltage pulse $U_0$ and radius $R_0$. At zero magnetic field, the screened  potential $U(r)$ is  modeled by \cite{Orlof,  SYLi,  Wyatt}:
\begin{figure}[tp]
\centering
\includegraphics[width=9cm, height=7cm]{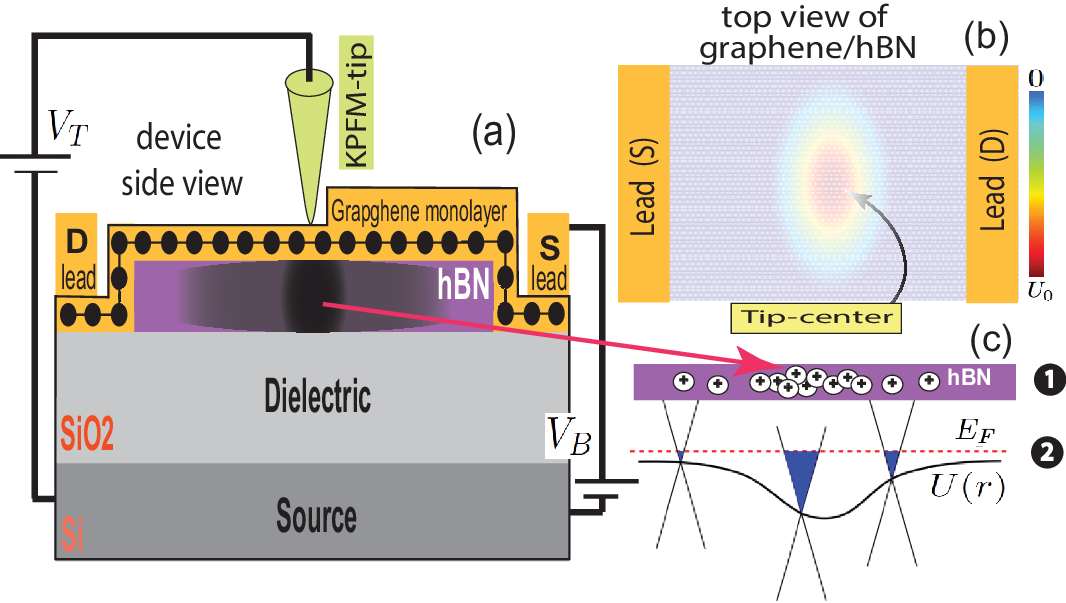}
\vspace{-0.45cm}
\caption{Schematic of a 
device to operate valley filtering. In (a) we show the  device, placed on top of a hBN substrate,  made of a graphene sheet with zigzag boundaries. The insulating substrate defines the dielectric area such as SiO$_2$ and a back-gate substrate (source) as Si, for instance. In (b) we show the total screened potential felt by the graphene sheet due to the  stationary charge distribution 
in line with experimental works \cite{Wyatt}. Panel (c) illustrates (1) the stationary distribution of the charges in the hBN substrate, due to the KPFM tip, and underneath it. 
(2) The sketch of the induced potential $U(r)$ and the Fermi energy $E_F$ where we illustrate the position of
	the Dirac point for a given position $r$
}
\label{fig1}
\end{figure}
\begin{eqnarray}\label{eq1}
U(r_i)\simeq  U_0  \exp\left( -
r_i^2/R_0^2 \right)+U_{\infty},
\end{eqnarray}
where  $r_i$ is the discretized distance of the graphene sites $i$ from the center of the KPFM tip. We denote by $U_0$ the electric potential at the center and $R_0 $ its corresponding radius. 
The third term $U_{\infty}$ defines the background value and can be controlled (cancelled out) by a back-gate voltage \cite{SYLi,  Wyatt}.

The model potential $U$, in Eq. (\ref{eq1}), is used in a 
tight-binding Hamiltonian to investigate the valley transport properties in the presence of the tip-induced potential. 
We adopt a tight-binding model in a honeycomb lattice holding a single $p_z$ orbital per site and neglect the chemical bonding or any modification in the atomic structure of graphene and hBN layers \cite{Marmolejo, GGiovannetti}. The resulting Hamiltonian that describes the system is given by 
\begin{eqnarray}
\label{eq2}
\notag
H = && -t\sum_{\langle i,j\rangle}{\bf a}_{i}^{\dagger} {\bf b}_{j} + \sum_{ \left\langle i \right\rangle }\Delta_{SG}\left(  {\bf a}_{i}^{\dagger} {\bf a}_{i}- {\bf b}_{i}^{\dagger} {\bf b}_{i} \right) \\
&& \qquad \qquad \quad  \qquad 
+\sum_{ \left\langle i \right\rangle } U_i \left(  {\bf a}_{i}^{\dagger} {\bf a}_{i}+ {\bf b}_{i}^\dagger {\bf b}_{i} \right).
\end{eqnarray}
where ${\bf a}_{i}^\dag$ $({\bf b}_{i}^\dag)$ and ${\bf a}_{j}$ $({\bf b}_{j})$  are the creation and annihilation operators for an electron in graphene sublattice A (B) at sites $i$ and $j$, respectively. The hopping energy is denoted by $t$ and the on-site term is set to zero (Fermi level). The 
heterostructure introduces an additional second term $\Delta_{SG}$ which describes the induced sublattice gap that arises mainly from the presence of the hBN substrate beneath the graphene layer \cite{Marmolejo, GGiovannetti}.

Theory and experiments have been analyzed and compared in the presence of a STM or KPFM tip, and have shown that the screened potential $U(r)$ depends on the radius $R_0$ of the KPFM tip. For a 
Gaussian shape they have used the range  $20$ nm $<R_0\le70$ nm \cite{PAMaksym, JLee, MFreitag, Grushevskaya}. 
Similarly, we will consider a graphene/hBN channel with zigzag edges, width $W=110$ nm, and length $L=300$ nm. Further, we take 
$t=-2.7$ eV and  $\Delta_{SG} = 29.26$ meV \cite{GGiovannetti, WYao, YRen},  a 
tip  radius $R_0=55$ nm, and  
$U_{\infty}=0$ since the  value of $U_{\infty}$ can be controlled 
 by a gate voltage \cite{Wyatt, SYLi}.

\begin{figure}[tp]
\centering
\includegraphics[width=9cm, height=6cm]{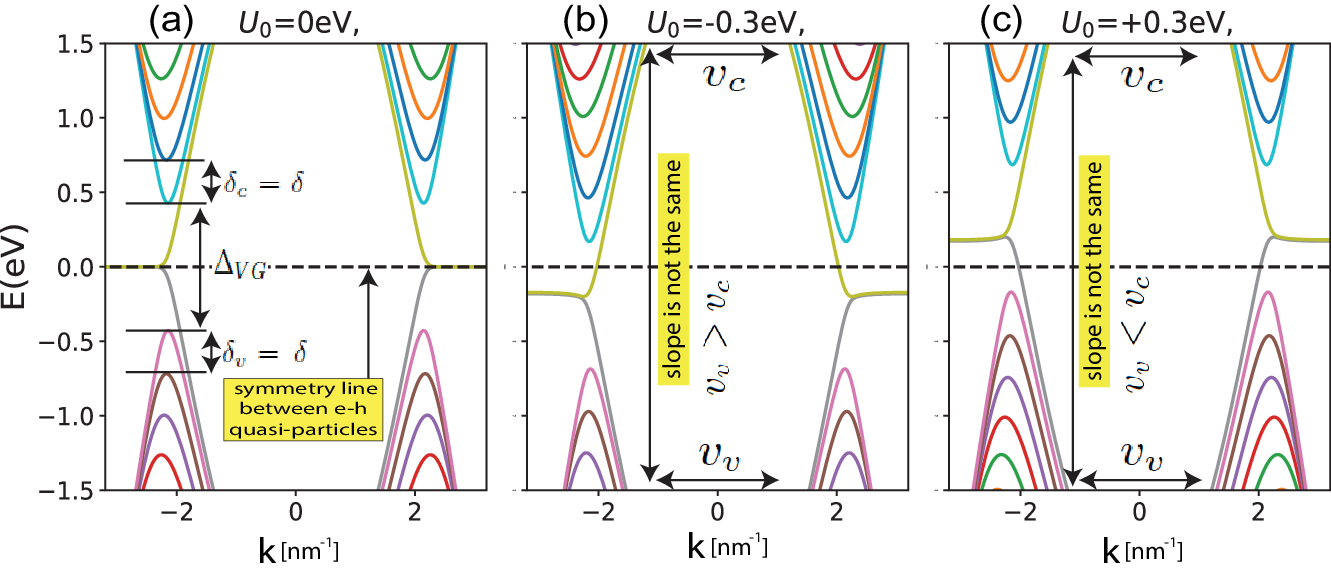}
\vspace{-0.45cm}
\caption{
Energy spectrum of a two-dimensional (2D) zigzaged strip  $3$ nm wide. 
Panels (a) shows the  spectrum
 in the absence of a tip-induced potential ($U_0 = 0$). When the Fermi energy $E_F$ is larger than $\Delta_{VG}/2$, where $\Delta_{VG}$ is the  valley-mode spacing gap, 
 the ${\bf -K}$ and ${\bf +K}$ channels are propagating. The valence and conduction mode spacings are denoted by
$\delta_v$ and $\delta_c$, respectively;   
Panels (b) and (c) show the spectrum in the presence of an induced potential with $U_0 = -0.3$ eV and $U_0 = +0.3$ eV, respectively. 
To see the effect of the broken e-h symmetry and compare  (a), (b) and (c), we keep the dashed line at the reference energy $E_F=0$.  We indicate the difference in 
slope of  the valence and conduction bands depending on the sign of $U_0$ for a given 
$E_F$: 
we have $v_v< v_c$ ($U_0<0$ ) or $v_v> v_c$ ($U_0>0$).}
\label{fig2}
\end{figure}

\vspace{-0.15cm}

\section{ Results and discussion}\label{sec3}
Below 
we   discuss how the Fermi energy $E_F$ and the induced KPFM potential lead to 
  valley filtering when only one valley channel is active and 
some conditions are fulfilled/ 
  We compute the transmittance of each valley 
  and show that the relevant conditions concern mainly the signs of $E_F$ and  $U_0$ and their  ranges. 
  
\subsection{ Electron-hole 
symmetry broken by the KPFM tip potential}\label{sec3.A}

Before stepping into the 
process of valley filtering and investigating the parameters that 
affect and monitor the valley transport in the presence of the induced electrostatic potential, we start by showing the dispersion relation for zigzag boundaries of the honeycomb lattice in Fig. \ref{fig2}. 
For operating valley filtering, it is important to have propagating modes at 
 both valleys. 
This 
is achievable in a 2D zigzag strip, when 
$E_F$ is 
higher than 
 $ |\Delta_{VG}/2|$,  where $\Delta V_G$ is the valley-mode spacing gap, as shown in Fig. \ref{fig2} (a). For this reason, the valley-dependent conductances in the system can be addressed independently only beyond this limit defined by what we call the valley-mode 
gap $\Delta_{VG}$ with both the ${\bf -K}$ and ${\bf +K}$ channels  propagating.

Additionally, it is clearly observed from the band spectrum in Fig. \ref{fig2} (b) and (c) that the electron-hole symmetry is broken by the induced potential, where the conduction and valence bands are not affected simultaneously by the same value of induced potential $U_0$  (non-vanishing value for the potential at the border of the system due to the finite size of the system). In fact, positive values of the induced potential affect the quasi-particles for $E_F>0$ while the negative ones affect them only for 
$E_F<0$. This broken symmetry between the quasi-bound states in the valence and conduction bands  
creates 
the correct conditions  for operating valley filtering of the propagating carriers at a given ${\bf k}$ 
and the Fermi velocity  plays a major role in selecting valley current as  will be illustrated  below. 

To  resolve different transport channels in ${\bf k}$-space, where the electron transmission probability of each Dirac cone is observed separately, we adopt the tight-binding model in Eq. \ref{eq1}, and we define the valley conductance $G _{-}$ and $G _{+}$ related to the current flow across the induced potential at given Dirac cones ${\bf -K}$ and ${\bf +K}$, respectively. More details about computing valley conductance are discussed in Appendix. \ref{app-A}. 

To investigate 
the dependence of the valley conductance in terms of the Fermi energy $E_F$ and tip-induced potential pulse $U_0$, we consider two cases: (1) the valley conductance is considered in terms of $E_F$ for a fixed value of $U_0$; (2) the valley conductance is considered in terms of $U_0$ for a fixed value of $E_F$.

\subsection{ Valley conductance 
in terms of the Fermi level}\label{sec3.B}
\begin{figure}[tp]
\includegraphics[width=9cm, height=4.8cm]{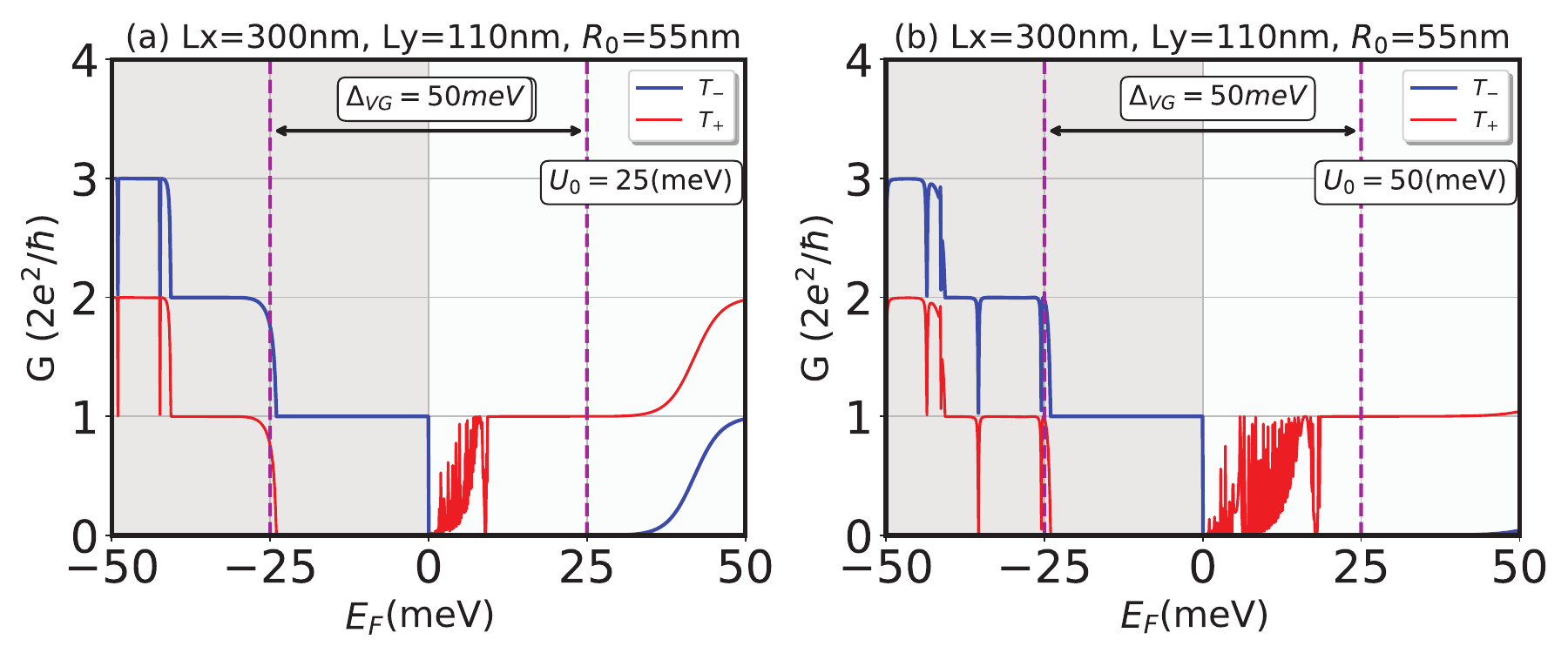}
\vspace{0.1cm}

\includegraphics[width=9cm, height=4.8cm]{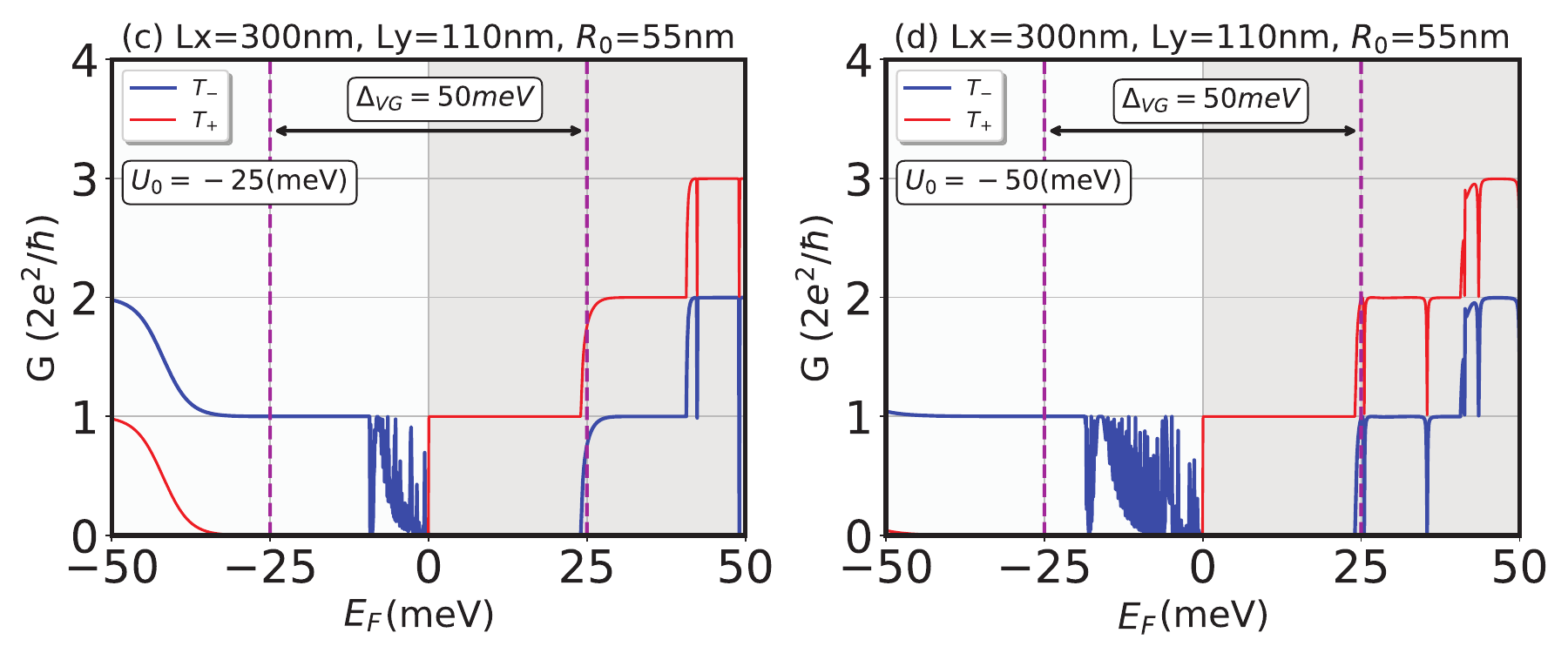}
\vspace{-0.55cm}
\caption{Valley conductance versus  Fermi energy for positive (top) and negative (bottom) induced potential $U_0$
as indicated. The red and dark blue curves show, respectively, the valley transmissions $T_+$ and $T_-$.
 $\Delta V_G$ is specified in the insets to all panels and the geometrical parameters on their tops.}
\label{fig3}
\end{figure}

We have calculated the valley-dependent transmission at  each valley 
 independently for  fixed  tip potential $U_0=\pm 25$ meV and $\pm 50$ meV.  The Fermi energy of the incident electrons varies between $-50$ meV and $+50$ meV and numerical results for the valley conductance, as a function of $E_F$, are shown in Fig. \ref{fig3}. 

It is clear that by tuning the Fermi energy $E_F$ one could operate a valley filter in a none symmetric energy range and within the first propagating mode defined by the energy mode $E_m^{(1)}$. Depending on the sign of the induced potential and within an energy range, only one valley channel is allowed to pass.  According to Fig. \ref{fig3},  the valley filtering happens when 
$E_F$ is increased beyond the energy limit 
$\Delta_{VG}/2$. We find that, for positive values of the induced potential, as shown in Fig. \ref{fig3} (a) and (b), only one valley is allowed within the energy range $\Delta_{VG}/2<E_F<E_m^{(1)}$ meV with $E_m^{(1)}=30$ meV ($50$ meV) for $U_0=25$ meV ($50$ meV). We observe that $100\% (0\%)$ of the conductance results from the flow of electrons at ${\bf +K}$ (${\bf -K}$) for positive $E_F$, 
while $50\%$ of the conductance results from the flow of electrons at both valleys for negative 
$E_F<-\Delta_{VG}/2$. 

The presence of the electrostatic potential induced from the KPFM tip does affect the quasi-bound states in the valence or conduction bands depending on the bias sign, as shown in Fig. \ref{fig2} (b) and (c). Consequently, the propagating modes of the electron quasi-particles (at positive energy) and hole quasi-particles (at negative energy) behave differently. As a result,  valley-dependent transport, when the electron-hole symmetry is broken , will depend on the sign of  $E_F$ and $U_0$. 
For instance, for $U_0>0$ ,  at positive $E_F$ 
the propagating modes are affected by the induced potential, and  valley-dependent transmission occurs 
$E_F>\Delta_{VG}/2$. However, at negative $E_F$ 
the propagating modes shift 
from the conduction to the valence bands, as shown in Fig. \ref{fig2} (c). This interband transition is not affected by the induced potential 
when $U_0$ is positive, and hence no 
valley-dependent  transmission occurs 
at $E_F< -\Delta_{VG}/2$.  Similarly, for negative induced potential, the electron propagating modes belonging to the conduction bands 
are not affected while the quasi-bound states in the valence bands are. Summarizing, 
valley filtering happens at $E_F<- \Delta_{VG}/2$ and destroyed at  $E_F>\Delta_{VG}/2$. 

On the basis of the above arguments, a valley-dependent  transmission, i.e., a 
selective population of a single valley, is pronounced depending on the sign of $U_0$ and $E_F$. In more detail, we contrast 
the slopes in the dispersion relations of 
the valence and conduction bands 
according to the  signs of $ E_F$ 
and the induced potential. This contrast 
does explain the presence (absence) of valley filtering at $E_F> |\Delta_{VG}/2|$ for positive (negative) values of $U_0$. More precisely, for a given  $ E_F$  
this slope does affect the Fermi velocities 
depending on the sign of $U_0$ as  shown in Fig. \ref{fig2}.  To confirm this assertion, we refer again to the dispersion relation, for zigzag boundaries,  where we express the  Fermi velocities in terms of the mode spacing $\delta$ or valley-mode 
gap $\Delta_{VG}$  as
%
\begin{equation}\label{eq3}
\Delta_{VG}= 
\sqrt{3} \pi t a/2W= 
\pi \hbar v/W
\end{equation} 
where $ v = (\sqrt{3}/2)t a/\hbar = 3\times 10^{6}$ m/s is the Fermi 
velocity in pristine graphene. In our case, with 
the tight-binding parameters and sample shapes specified in Sec. \ref{sec2}, we have $\Delta_{VG} =50$ meV,
where $\delta_v=\delta_c=\Delta_{VG}/3=  \pi \hbar v /3W$. Hence, the mode spacing is straightforwardly derived from the  velocity and vice versa.

One important remark that we might also highlight from the output of Fig. \ref{fig3} is the presence of an oscillatory behavior that is valley dependent. Indeed, the oscillations near the vicinity of the mode opening energy is appearing due to the potential in the scattering region, where only few-mode (valley-dependence) are affected by the potential landscape of the GQDs and behave similarly to the Fabry-Perot oscillation \cite{Goldman}. 
More precisely, the conductance oscillations are valley-dependent and many features of conductance oscillations are similar to the resonant tunneling in quantum antidots \cite{Mills, Camino}. Importantly, as shown in Fig. \ref{fig3}, ${\bf K_1}$ (${\bf K_2}$) valley modes are affected by tip-potential landscape and feature conductance oscillations at negative (positive) Fermi levels where the resonance increases proportionally with induced potential and happens only within the valley-mode gap when both $E_F$ and $U_0$ have the same polarity.

\subsection{ Valley conductance in terms of the induced potential}\label{sec3.B}

When a tip potential is induced, the Fermi velocity, at positive or negative incident energy around the tip-induced potential is no longer the same 
and becomes a function of $U_0$.
 Indeed, The presence of the contacts (left and right reservoirs) makes the system finite and therefore the tip's induced potential doen't vanish near the leads. One has to consider that remanescent component of the potential in the lead and thus ends up with a band structure  with different Fermi velocities ($v_c$, $v_v$) at the conduction and valence bands.
 
We bear in mind that the Fermi wavelength $\lambda_F$ is inversely proportional to the Fermi velocity $v_F$.  
The conductance is very sensitive to the variation of $\lambda_F$ (especially for large quantum dots). In fact, far from the modes opening, the Fermi velocity approaches that of infinite pristine graphene and therefore it barely varies. In contrast, near the mode opening, where the band is highly non-linear, the Fermi velocity  varies a lot with $E_F$ and this explains why depending on the sign of ($U_0\times E_F$,) $\lambda_F^{-K}\ne\lambda_F^{+K}$ and therefore, as we can deduce from the band structure, the filtering can happen or not.

Now let us go back to Fig. \ref{fig3} and discuss the range of valley filtering. It is seen that by increasing the value of the induced potential $U_0$ from $25$ meV to  $50$ meV, the energy range of the valley filtering increases since the energy mode 
is sensitive to the value of $U_0$ 
($E_m^{(1)} \propto U_0$) and steps from $30$ meV to $50$ meV, respectively. From Fig. \ref{fig2} and Fig. \ref{fig3}, we 
infer that for positive $U_0$ and 
$E_F$, the electron propagating modes 
are strongly affected and the energy mode $E_m^{(1)}(U_0=25$ meV) $\neq E_m^{(1)}(U_0=50$ meV) where the conductance exhibits a smooth less quantized plateaus. For  negative $E_F$ the modes are not affected by $U_0$ where $E_m^{(1)}(U_0=25$ meV)$=E_m^{(1)}(U_0=50$ meV), and hence the conductance exhibits quantized plateaux at an odd number 
of $2e^2/ \hbar$ where $2$ stands for spin degeneracy. However, for negative induced potential, the process is entirely inverted because for  negative $E_F$ the propagating modes  are strongly affected whereas for positive $E_F$ they are not.

\subsection{Rules for operating selective valley current}\label{sec3.C}
The analysis of Sec. \ref{sec3.A} and \ref{sec3.B} showed that  valley filtering is allowed only when the sign of the product ($E_F \times U_0$) is positive. Indeed, beyond the valley-mode spacing gap ($E_F>\Delta_{VG}$) in  Fig. \ref{fig3}, we showed that a positive  (light gray background)  product leads to valley filtering of the current while a negative one (dark-gray background) destroys the valley filtering process. 
\begin{figure*}[tp]
\centering
\includegraphics[width=14cm, height=10cm]{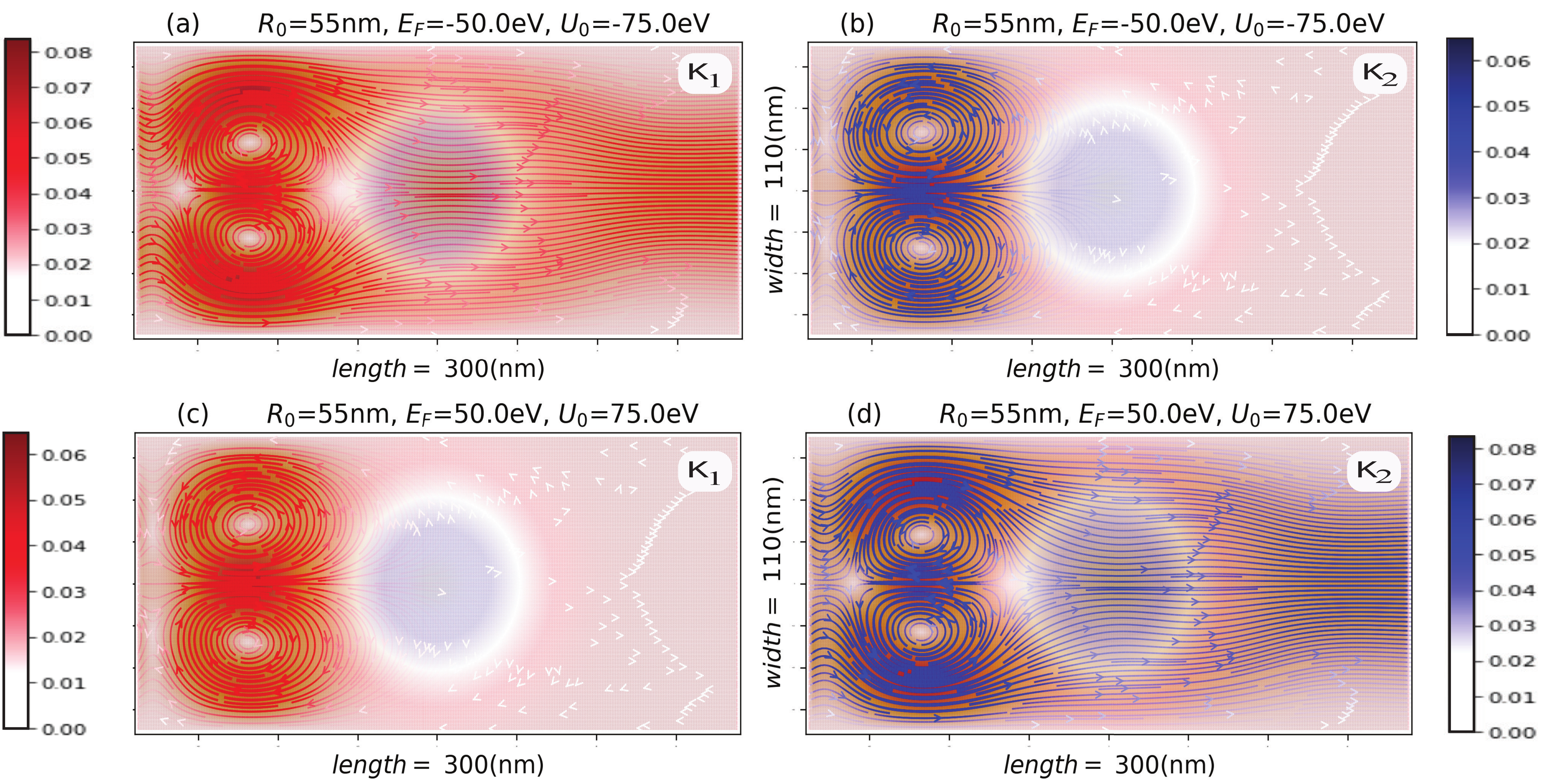}
\vspace{-0.35cm}
\caption{Real-space mapping of both valley currents (red (blue) lines show the ${\bf K_1=-K}$ (${\bf K_2=+K}$) current)  in the presence of the induced potential $U_0$. Panels (a) and (b) are for negative energy, while (c) and (d) are for positive energy. The sign of $U_0$ is set the same as that of $E_F$. The induced potential of the KPFM tip traps the charge in the hBN space and its effect (based on the Poisson equation) is illustrated by the spherical shape in the middle of the graphene sample (see map in Fig. \ref{fig1} (b)) held a few nanometers from the surface of graphene;  it decays to zero for $ 
(x^2+y^2)^{1/2}>R_0$.}\label{fig4}
\end{figure*}
This  is also shown in Fig. \ref{fig4}, where we plot the valley current in terms of $E_F$ and  $U_0$ and maintain the product $E_F\times U_0$ positive. 

In more detail, we set the sign of $U_0$ the same as $E_F$ and then select a positive (negative) value of $E_F$ 
between $\Delta_{VG}/2 $ $ (-\Delta_{VG}/2)$ and $E_m^{(1)} (-E_m^{(1)})$, cf. Fig. \ref{fig3}. Once these conditions are fulfilled, we map the current of the propagating channel at ${\bf {\bf K_1=-K}}$ and ${\bf {\bf K_2=+K}}$. 
The corresponding current is 
evaluated and 
mapped in Fig. \ref{fig4}.

As illustrated in Fig. \ref{fig4}, the valley filtering process is operative due to the positive sign of the product $E_F \times U_0$.  Interestingly, the currents  for both positive and negative $E_F$ are equal, but the opposite energy sign shifts the valleys with only one valley allowing current  to flow and the other one  blocking it.  Hence, depending on $E_F$ and $U_0$, one can break the valley degeneracy and generate a valley-polarized current. This is an important result as it leads to 
valley selection by changing either $U_0$ 
or a bias gate which shifts $E_F$ or changes its sign. 

Below we will show that the valley filtering can also take place for some  potentials and either sign of the product 
 $E_F\times U_0$. This valley filtering does correspond to  resonances with some states affected by the induced potential. 
\subsection{Valley filtering and resonances}\label{sec3.D}
\begin{figure}[tp]
\includegraphics[width=9cm, height=6cm]{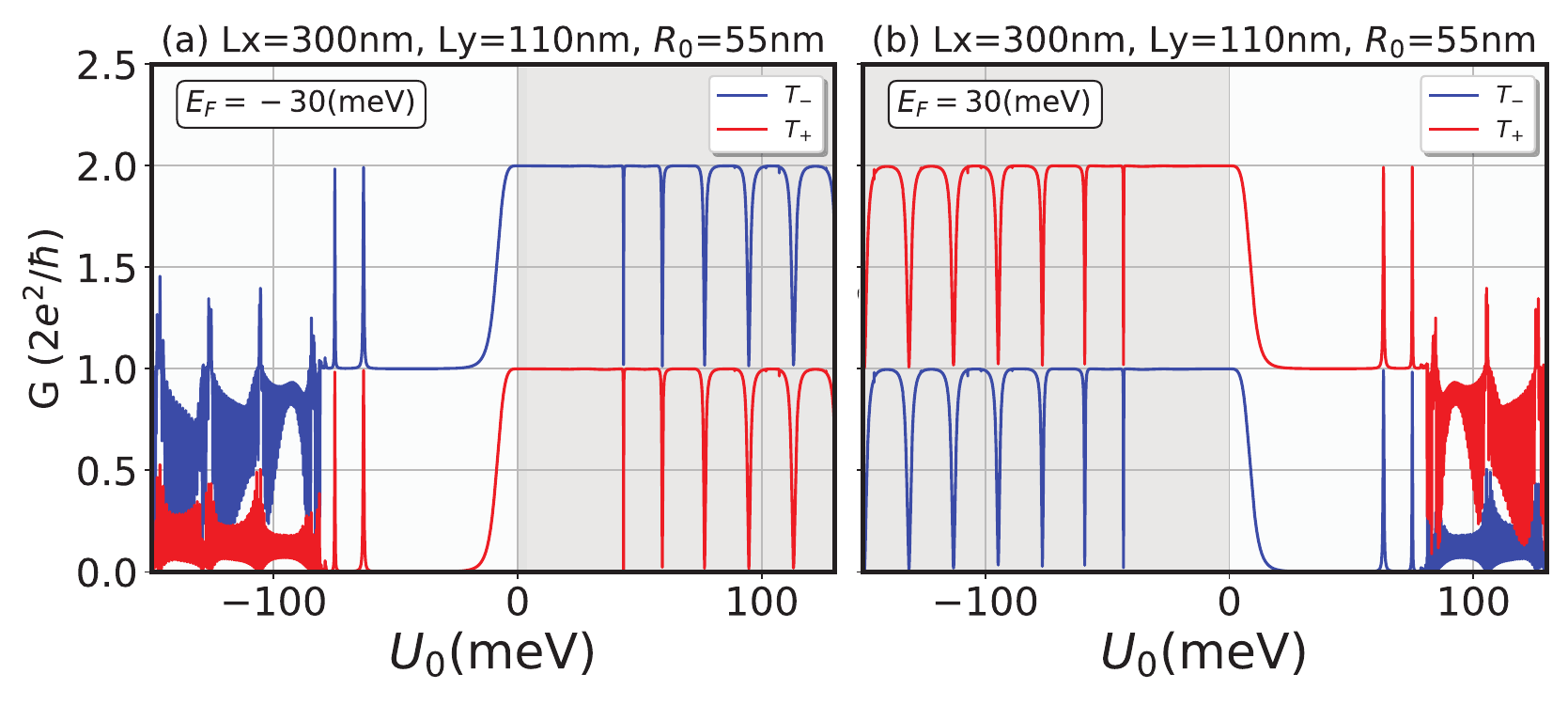}
\includegraphics[width=9.5cm, height=8cm]{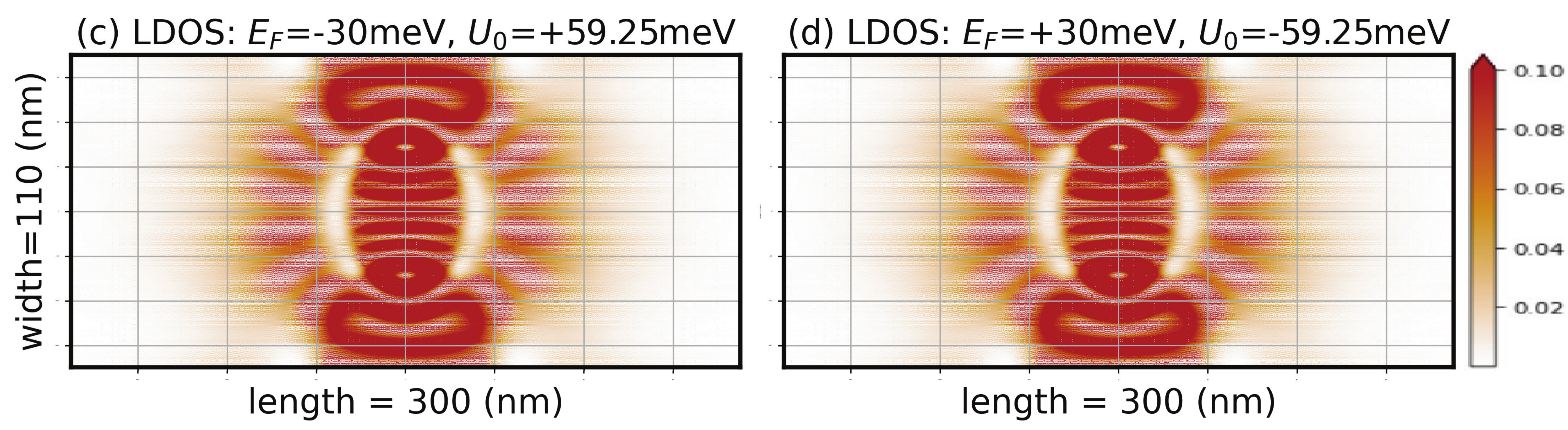}
\vspace{-0.35cm}
\caption{Valley conductance versus  tip-induced potential for negative (a) positive (b) Fermi level $E_F$. Panels (c) and (d) show the corresponding local density of  resonant states. 
$U_0$ and $E_F$ have opposite signs.}
\label{fig5}
\end{figure}

As mentioned above, the tip-induced potential  $U(r)$ breaks the symmetry between the valence ($E_F<0$) and conduction ($E_F>0$) bands and the valley-polarized conduction becomes sensitive to its sign and strength 
for a given $E_F>0$. 
 In Fig. \ref{fig5} (a) and (b) we show the conductance as a function of  $U_0$.  

First, as in section \ref{sec3.B}, the results show that the 
valley filtering depends on the sign of the product $E_F \times U_0$ which is the key point for breaking the valley degeneracy and creating a valley-polarized current. 
For instance, for $E_F \times U_0>0$ and at  $E_F=-30$ meV, the  conductance is polarized for $U_0$ in the 
range $-60$ meV $<U_0<-\Delta_{VG}/2$ with only the ${\bf K_1=-K}$ channel  conducting. However, for $E_F=+30$ meV  the 
valley-selecting process is reversed  and only the ${\bf K_2=+K}$ channel is conducting for $U_0$ in the 
range $-\Delta_{VG}/2<U_0<+60$ meV. 

Second, for $E_F \times U_0<0$ the obtained results show 
valley anti-resonances (and resonances) where the conductance drops to zero with respect to valley conductance at only one Dirac cone ($K_1$ or $K_2$). This process can be justified due to valley confinement, known as Klein's tunneling \cite{Wyatt, CGutierrez, Silvestrov}. More precisely, to show the presence of  confined states we employ the kernel polynomial method (KPM) 
to numerically compute the 
local density of states (LDOS) using Chebyshev polynomials \cite{aw2006, af2015} along with damping kernels \cite{Wellein} as 
recently provided by a Pybinding package \cite{md2017}.  To compute the LDOS 
we count the sites contained within the  shape 
of the induced potential, determined by $(x^2+y^2)^{1/2}<R_0$. 
We observe that the electron are almost localized at induced potential landscape, where the superposition of the confined  states wind up with the features of vortex pattern which does appear at the induced potential boundaries.

The same remarks as in Fig. \ref{fig4}, can be drawn from  Fig. \ref{fig5}(c) and (d) where the LDOS for ${\bf K_1=-K}$ and ${\bf K_2=+K}$ are equal, where the opposite energy sign shifts only the valleys with only one valley  confined. Hence, depending on $E_F$ and $U_0$, one can break the valley degeneracy and generate valley confined states when the  product $E_F \times U_0$ is  negative.

The resonance at $U_0=+59.25$meV ($U_0=-59.25$meV) occurs 
for negative (positive) 
$E_F$ and results from 
confined states of the quasi bands in the valence (conduction) band. 

The main point here is that we confirm and show that the states in the case $U_0=+59.25$meV ($U_0=-59.25$meV) are indeed resonant states with a high local density within the area that defines the GQDs.
In our case, the interference might happen inside the induced island due to the shape of GQDs with specific values of induced potential.

Since we are dealing with electron-hole broken symmetry, 
positive and negative energy bands are affected independently by the induced potential. Also, 
since the ribbon width $W$ is finite,  the momentum is discretized. 
Therefore, 
 the anti-resonances for 
$E_F \times U_0<0$ (dark gray area in Fig. \ref{fig5} (a) and (b)), can be clearly identified and appear 
nearly periodic. Since $E_F$ does affect the set of discrete values in  momentum space,  we can state that different values of 
$E_F$ lead to a different set of resonances  
with their number 
depending on 
the values of $E_F$ and  $W$.
\subsection{Robustness of 
valley filtering against disorder  and strip width} 
\label{sec3.E}
\begin{figure}[tp]
\includegraphics[width=9cm, height=6cm]{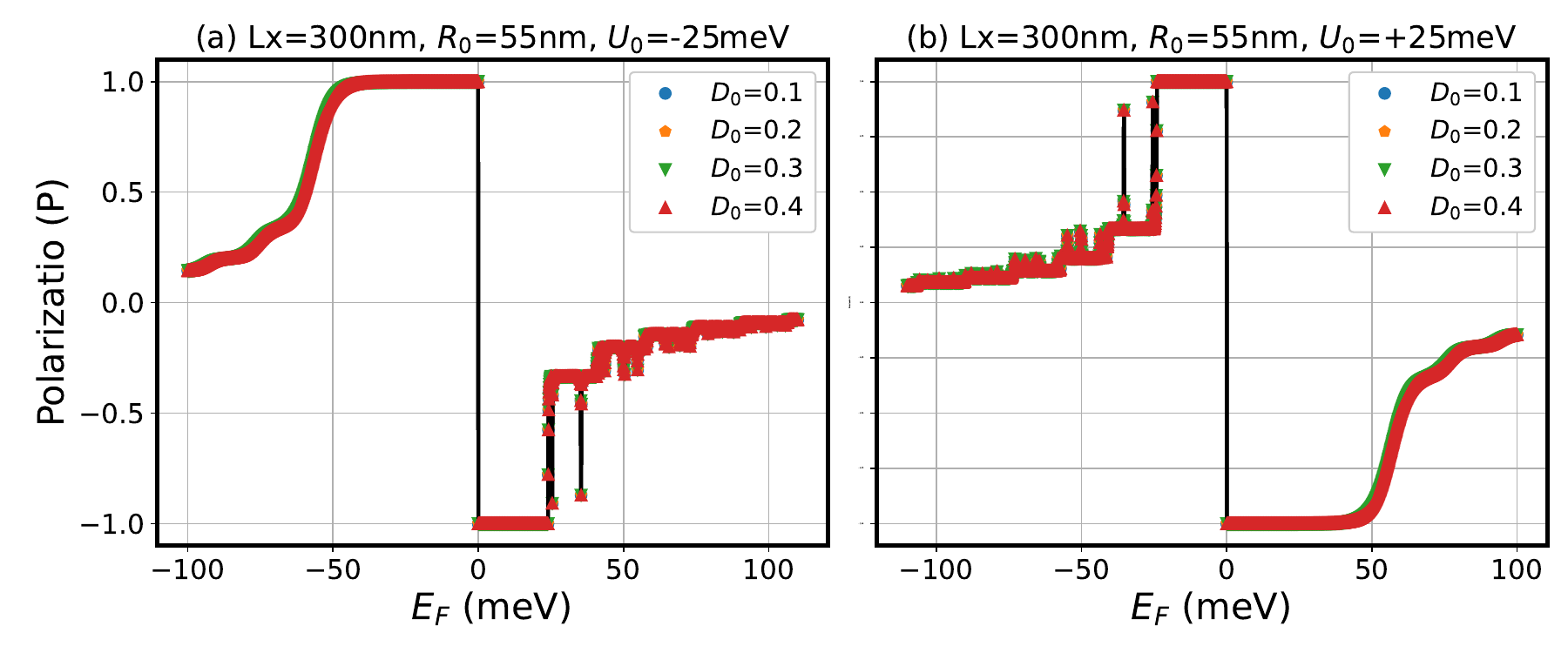}
\vspace{-0.85cm}
\caption{Polarization versus  Fermi energy for (a) negative and  (b) positive  induced potential.}
\label{fig6}
\end{figure}

Operating valley filtering controlled by either $U_0$ 
or $E_F$ 
must be robust 
against a disorder potential. For this purpose, in Fig. \ref{fig6}, the valley polarization is plotted as a function of $E_F$ 
in the presence of an on-site disorder of 
strengths $D_i$. The relevant 
Hamiltonian is $H_D = H+ \sum_{ \left\langle i \right\rangle } D_i \left(  {\bf a}_{i}^{\dagger} {\bf a}_{i}+ {\bf b}_{i}^\dagger {\bf b}_{i} \right)$, where $H$ is defined in Eq. (\ref{eq2}) and $D_i$ are numbers randomly 
distributed in the range $[-D_0, +D_0]$. We will consider a strong disorder 
$5U_0<D_0<15U_0$.
\begin{figure}[tp]
\includegraphics[width=9cm, height=6cm]{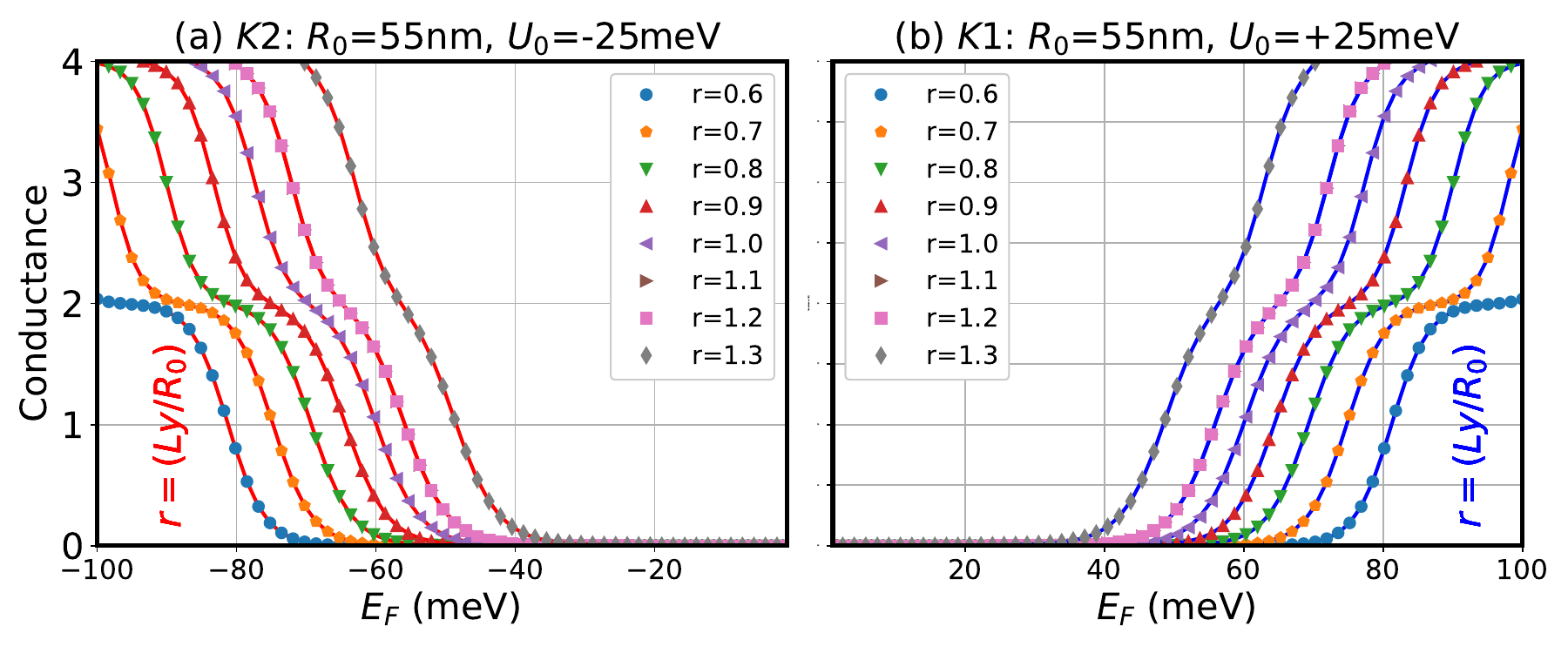}
\includegraphics[width=9cm, height=6cm]{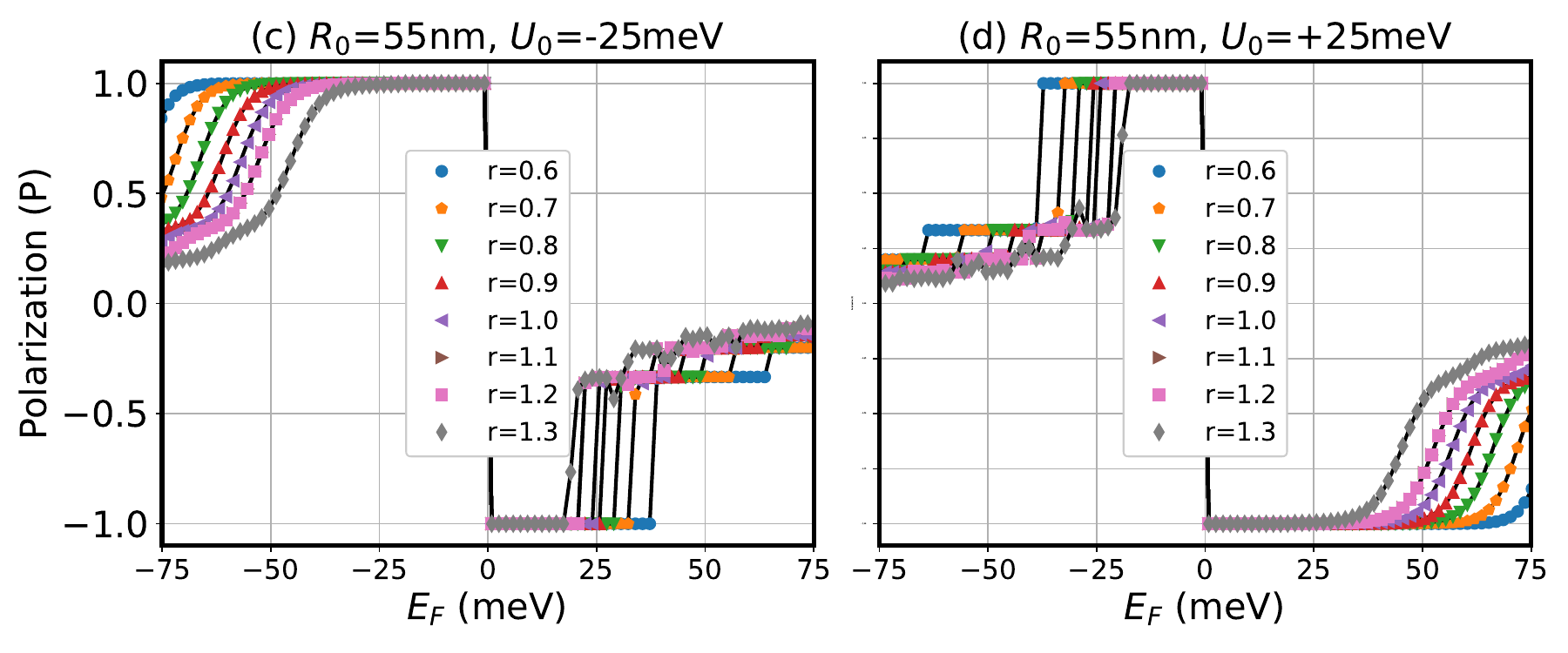}
\vspace{-0.85cm}
\caption{The top panels show the valley conductance, at $K_2$ in (a) and at $K_1$ in (b), versus $E_F$ for several widths $W$
determined by $r = W/R_0$. The bottom panels show the polarization versus $E_F$ for several values  $r$.
}
\label{fig7}
\end{figure}
We notice that the disorder does not affect the polarization even for  values stronger than $D_0=15U_0$. For all considered  disorder strengths, Fig. \ref{fig6} shows that  a valley filtering is always present and 
robust against on-site disorder.  


Additionally, we have also considered the effect of the ribbon width and 
 plotted, 
  in Fig. \ref{fig7}, the valley conductance versus $E_F$ for several widths $W$, determined 
  by the ratio $r=W/R_0$, for a Gaussian shape with $R_0=35$ nm. 
We  focus on the side on which $E_F\times U_0>0$ and valley filtering operates as discussed previously. We notice 
 that the valley filtering is more  evident for $r \leq 1$. More precisely, for $r=0.6, 0.7$, $0.8$, we have, respectively, the energy ranges $-0.72\leq E\leq +0.72, -0.64\leq E\leq +0.64$, and $-0.58\leq E\leq +0.58$ meV, where the $+$($-$) signs are for $U_0=+25$ meV ($U_0=-25$ meV), respectively.  From Fig. \ref{fig7}(a) and (b), we can see that the conductance plateaux are flatter for $r<1$ 
than for $r>1$. Additionally, from Fig. \ref{fig7}(c) and (d) we clearly observe that by increasing the ratio $r$ the energy range of controlling the valley filtering decreases and it might vanish for $r>1$ since $E_F$ falls between $0$ and $\Delta_{VG}/2$. More precisely, for $r>1$ the polarization drops and we have 
$P<\left| 1 \right|$ for $E_F> \left| \Delta_{VG}/2 \right|$. 



\section{summary and conclusions}\label{sec4}
We presented an approach for operating valley filtering based on the KPFM-induced potential that opens various roads to experimental verification. Using such an electrostatic potential, instead of PMFs induced from nanobubbles, we can operate or destroy the valleys filtering depending on the signs of the electron energies and the induced potential.  
A positive sign of their product ($U_0 \times E_F>0$) allows operating valley filtering, and the bias voltage, which controls the energy sign, shifts the valleys while only one of them allowing current to flow  and the other one blocked. 

We have also noticed the presence of conductance oscillations near the vicinity the mode opening energy, which are valley dependent and whose strength is proportional to that of the  induced potential within the mode-spacing gap. 
These oscillations are similar to those in resonant tunneling in quantum antidots and to the Fabry-Perot ones.
Furthermore, valley polarized currents can occur for negative products $U_0 \times E_F<0$. In such a case 
the valley filtering does correspond to  resonances;  some states are affected by the induced potential and 
only propagating states belonging to one valley 
confined in the induced GQDs occur.

To the best of our knowledge, we are the first to realize that the valley is controlled and might be processed by the rule defined by the  sign of $U_0 \times E_F$, where the interplay between the sign of $E_F$ 
 and that of the tip-induced potential provides an alternative example of valley filtering. The results of the present study can facilitate the development of valleytronic devices. \\
 
{\bf Acknowledgments.}
 The authors acknowledge computing time on the SHAHHEN supercomputers at  KAUST University.


\clearpage
\begin{widetext}
\begin{appendix}
\numberwithin{equation}{section}
\section{Valley-dependent transmission and polarization}\label{app-A}
Below we briefly describe the derivation of the valley-dependent transmittance and polarization expressions. 
The propagating modes in the leads can be selected 
depending on their velocity and momentum direction. This is achieved using the 
Kwant functionalities \cite{kwant} that couple 
the propagating modes 
with the scattering region and therefore allow the evaluation of  valley transport properties. 
We consider only propagating modes and  assume that valley modes are defined based on propagating states $\Phi( {\bf  v<0})$ \cite{kwant}. These states are characterized by both degrees that contain the two valleys $K_1$ (obtained from $\Phi({\bf k}<0, {\bf v<0})$) and ${\bf K_2}$ (obtained from $\Phi({\bf k}>0, {\bf v<0})$) in the graphene lead \cite{Stegmann, settnes}. Once the valley states are defined, we resolve different transport channels in  reciprocal space, with the electron transmission probability at each Dirac cone computed  separately. Within the Green's function approach \cite{Istas, Ozaki} the valley-resolved channels lead to the total transmittance of electrons $T = T_{-K}+ T_{+K}$,  where the valley transmittance $T_{\pm K}$  given by 
\begin{equation}\label{eqA1}
T_{\pm {\bf K}}^{m, n}= \text{Trace} [G_{\bf \pm K} \Gamma^m G^\dagger_{\bf \pm K} \Gamma^n  ], \qquad (m, n = L, R);
\end{equation}
The Green function matrices are given by
\begin{equation}\label{eqA2}
G (\epsilon, {\bf \pm K}) = \left[ \left( \epsilon + \iota \eta \right)I-H^h({\bf \pm K})- \Sigma \right]^{-1} 
\end{equation}
and
\begin{equation}\label{eqA3}
\Gamma = \iota (\Sigma - \Sigma^{\dagger}).  
\end{equation}
$\Gamma$ is the imaginary part of the self-energy of the contact given by coupling, independently, the scattering region (defined by the Hamiltonian $H^h$ ) with each valley mode.  For more details see Ref. \cite{Stegmann, Stegmann22}.
Once the valley-dependent transmission is derived,  we define the valley conductance $G_-$ and $G_+$ at the Dirac cones  ${\bf -K}$ and ${\bf +K}$, respectively as  $G_{\pm} =   (e^2/h) T_{\pm {\bf K}}$
To obtain both valley modes and ensure valley-resolved channels, we consider the propagating modes for $E_F>abs(\Delta_{VG}/2)$ (cf, Fig. \ref {fig2}). After obtaining them 
the two valleys can be separated depending on their momentum sign. The resulting  
valley polarization is obtained as
\begin{equation}\label{eqA4}
P=\frac{T_{{\bf -K}}-T_{{\bf +K}}}{T_{{\bf -K}}+T_{{\bf +K}}}.
\end{equation}
For $P = \pm 1$ the electrons are localized entirely at the $\pm {\bf K}$ valley and full polarized transmittance is ensured.
$P = 0$ corresponds to unpolarized  electrons. 

We might also obtain the local density of states (LDOS) at a given sample site $i$ as:
\begin{equation}\label{eqA5}
\text{LDOS}\left( E \right)=\sum_{l}^{}\left| \left\langle i|l \right\rangle \right|^{2}\delta\left( E-E_l \right)
\end{equation}
the energy $E$ is the energy of the confined states where somation goes over all electron eigenstates $|l> = c_l^{\dagger}|0>$ of the Hamiltonian $H$ in Eq. \ref{eq2} with energy $E_l$.  The quantity in Eq. \ref{eqA5} is numerically computed  using Chebyshev polynomials \cite{aw2006, af2015} and damping kernels \cite{Wellein}. 

\section{\textit{Valley current mapping}}\label{app-B}
We adopt the procedure  detailed in the Kwant package \cite{kwant}. 
The density operator and continuity equation are expressed as
\begin{equation}\label{eqB1}
\rho_q=\sum_{a}\Phi^{*}_{a} H^h_{q} \ \Phi_{a}, \qquad  \qquad
\frac{\partial \rho_a}{\partial t}-\sum_{b}^{}J_{a, b}=0.
\end{equation}
$H^h$ is the Hamiltonian of the heterostructure in the scattering region whose size is $N_1 \times N_2$ sites and $\Phi({\bf v<0})$ is the  eigenstate of the propagating mode of the graphene's lead whose size is $N_1$. Here $q$ defines all sites or hoppings in the scattering region and $J_{ab}$ is the current.

For a given site of density  $\rho_a$, we sum over its neighbouring sites b.  
Then  the valley current $J_{\pm {\bf k}}^{ab}$ takes the form 
\begin{equation}\label{eqB2}
J_{{\bf -K}}^{ab}=\Phi^{*}({{\bf k}<0, {\bf v<0}})\left(i \sum_{\gamma}^{}H^{*h}_{ab\gamma} H^h_{a\gamma}-H^h_{a\gamma}H_{ab\gamma} \right) \Phi ({{\bf k}<0, , {\bf v<0}}),
\end{equation}
and 
\begin{equation}\label{eqB3}
J_{{\bf +K}}^{ab}=\Phi^{*}({{\bf k}>0, {\bf v<0}})\left(i \sum_{\gamma}^{}H^{*h}_{ab\gamma} H^h_{a\gamma}-H^h_{a\gamma}H_{ab\gamma} \right) \Phi^({{\bf k}>0, {\bf v<0}}),
\end{equation}
\noindent where $H_{ab}$ is a matrix with zero elements except for those connecting the sites a and b. In this case, the hopping matrices in the heterostructure are obtained from the first term of Eq. \ref{eq2}.
\end{appendix}
\end{widetext}

\textcolor{white}{text}

\end{document}